%% file: letter.tex
\documentclass[aps, prd, twocolumn, nofootinbib, floatfix, superscriptaddress]{revtex4-1}

\usepackage{comment}
\usepackage{graphicx}
\usepackage{dcolumn}
\usepackage{bm}
\usepackage{subfigure}
\usepackage{epsfig}
\usepackage{amsmath}
\usepackage{amsfonts}
\usepackage{graphicx}
\usepackage{amssymb}
\usepackage{nccmath}
\usepackage{amssymb,amsmath,amsfonts}
\usepackage{xfrac}
\usepackage{color}
\usepackage{tikz}
\usepackage[T1]{fontenc}
\usepackage[utf8]{inputenc}
\usepackage{ulem}

\usepackage{tcolorbox}
\usepackage{hyperref}
\usepackage{booktabs}
\hypersetup{colorlinks=true, citecolor=red, linkcolor=blue, urlcolor=blue}

\definecolor{rossos}{cmyk}{0,1,1,0.55}
\definecolor{bluscuro}{rgb}{0.15, 0.2, .85}
\definecolor{bluchiaro}{cmyk}{1,.3,0.,0.1}

\input{newcommands}

\newcommand{\be}{\begin{equation}}

\begin{document}
\rightline{}
\title{Is inflationary magnetogenesis sensitive to the post-inflationary history?}



\author{Konstantinos Dimopoulos}
\email{k.dimopoulos1@lancaster.ac.uk}
\affiliation{Consortium for Fundamental Physics, Physics Department,
Lancaster University, Lancaster LA1 4YB, United Kingdom}

\author{Anish Ghoshal}
\email{anish.ghoshal@fuw.edu.pl}
\affiliation{Institute of Theoretical Physics, Faculty of Physics, University of Warsaw, \\ ul. Pasteura 5, 02-093 Warsaw, Poland}

\author{Theodoros Papanikolaou}
\email{t.papanikolaou@ssmeridionale.it}
\affiliation{Scuola Superiore Meridionale, Largo San Marcellino 10, 80138 Napoli, Italy}
\affiliation{Istituto Nazionale di Fisica Nucleare (INFN), Sezione di Napoli, Via Cinthia 21, 80126 Napoli, Italy}
\affiliation{National Observatory of Athens, Lofos Nymfon, 11852 Athens, 
Greece}



\begin{abstract}
Considering inflationary magnetogenesis induced by time-dependent kinetic and axial couplings of a massless Abelian vector boson field breaking the conformal invariance we show in this article that, surprisingly, the spectral shape of the large-scale primordial magnetic field power spectrum is insensitive to the post-inflationary history, namely the barotropic parameter ($w$) and the gauge coupling functions of the post-inflationary era.
\end{abstract}


\maketitle

\vspace{-0.3cm}
\section{Introduction}
\vspace{-0.3cm}
Cosmological magnetic fields are observed at intergalactic scales of $\sim 4 {\rm Mpc}$ with field strengths ranging between $10^{-17}$ and $10^{-14}$ \cite{Ando:2010rb,Tavecchio_2010,Neronov_2010,Tavecchio_2011,Essey:2010nd,Finke:2015ona}. However, their origin remains a mystery to date, and constituting an active field of research. In particular, various primordial magnetic field generation mechanisms have been proposed such as inflationary ones~
\cite{PhysRevD.37.2743,1992ApJ...391L...1R,PhysRevD.46.5346,Dolgov:1993vg,Davis:2000zp,Dimopoulos:2001wx,Gasperini:1995dh,Martin:2007ue,Demozzi:2009fu,Kanno:2009ei,Emami:2009vd,Bamba:2003av,Bamba:2006ga,Kobayashi:2014sga,Barnaby:2012tk,BazrafshanMoghaddam:2017zgx,Durrer:2022emo,Kushwaha:2020nfa,Ferreira:2013sqa,Ferreira:2014hma,Ng:2014lyb,Tripathy:2022iev,Tripathy:2024ngu,Kushwaha:2022bwy}, phase transitions~\cite{1988PhRvD..37.2743T,1989ApJ...344L..49Q}, cosmic strings \cite{Dimopoulos:1997df,Davis:2005ih,Battefeld:2007qn}, primordial scalar/vector perturbations~\cite{Ichiki:2006cd,Naoz:2013wla,Flitter:2023xql,Banerjee:2003xk,Durrer:2006pc} as well as astrophysical processes where magnetic fields are seeded by battery-induced mechanisms~\cite{1950ZNatA...5...65B,Widrow:2002ud,Hanayama:2005hd,Safarzadeh:2017mdy,Araya:2020tds,Papanikolaou:2023nkx,Papanikolaou:2023cku}.

In the present article, we focus on inflationary magnetogenesis mechanisms of primordial origin, where it is well known that one needs to break the conformal invariance in the gauge sector (see Refs.~\cite{Kandus:2010nw,Subramanian:2009fu} for reviews on this topic) in order to generate very large magnetic fields at galactic/intergalactic scales. One natural source of breaking of such conformal invariance is due to the presence of inflaton-gauge field couplings, gauge field-curvature couplings \cite{Mazzitelli:1995mp,PhysRevD.37.2743} or the presence of additional spectator fields during inflation\cite{Giovannini:2007rh,Patel:2019isj,Giovannini:2021thf}. In particular, the very well-studied Ratra models~\cite{Ratra:1991bn} (see Refs.~\cite{Talebian:2021dfq,Talebian:2020drj,BazrafshanMoghaddam:2017zgx} for recent works on this topic) introduce a non-minimal coupling of the form $f(\phi)^2 F_{\mu\nu}F^{\mu\nu}$, where $\phi$ is the inflaton field and $F_{\mu\nu}$ is the electromagnetic field strength tensor. 

In the vanilla $\Lambda$CDM standard model of cosmology, it is very frequently assumed that the cosmic inflation era is followed immediately by the radiation-dominated epoch of the Hot Big Bang Universe. However, as we will highlight, the Universe can only become radiation-dominated at the end of inflation only under very restrictive assumptions in a simplistic framework. Indeed, in order to release all of its energy density right after it completes and exits the phase of slow roll, the inflaton must decay immediately into ordinary radiation bath. Such a fast decay of the inflaton field inevitably requires large interaction terms between the inflaton field and Standard Model (SM) fields. Notably, sizable interactions of the inflationary sector were shown to substantially affect any inflationary dynamics during this stage\cite{Buchmuller:2014pla,Buchmuller:2015oma,Argurio:2017joe,Heurtier:2019eou} or the stability of the SM Higgs boson \cite{Enqvist:2016mqj, Kost:2021rbi} spoiling as well the required flatness of the inflaton potential and/or giving rise to non-Gaussianities which so far remain unobserved in cosmic microwave background (CMB). 

However, several inflationary scenarios that involve runaway scalar field potentials do not actually naturally lead to a close or finite minimum around which the scalar inflaton may efficiently decay to produce the radiation bath particles. This is, for instance, the case encountered in quintessential inflationary scenarios~\cite{Peebles:1998qn, Dimopoulos:2001ix,Dimopoulos:2002hm,Dimopoulos:2017zvq, Dimopoulos:2017tud,Rubio:2017gty,Geng:2017mic,DeHaro:2017abf,Bettoni:2018pbl,Haro:2018zdb,Dimopoulos:2019gpz,Dimopoulos:2020pas,Dimopoulos:2022rdp,Akrami:2017cir,Bettoni:2021qfs,deHaro:2021swo} or, more generally, non-oscillatory inflation models \cite{Ellis:2020krl}. Here, the inflation sector thus only transfers a fraction of its energy to SM particles, however efficient reheating in those scenarios occurs via other alternative reheating mechanisms like gravitational reheating~\cite{Ford:1986sy,Chun:2009yu,Hashiba:2018tbu,Haque:2022kez}, instant preheating~\cite{Felder:1998vq,Campos:2002yk,Dimopoulos:2017tud}, curvaton reheating~\cite{Feng:2002nb,BuenoSanchez:2007jxm}, primordial black hole (PBH) induced reheating~\cite{Lennon:2017tqq,Martin:2019nuw,Dalianis:2021dbs}\footnote{See Refs. \cite{Bhaumik:2022pil,Bhaumik:2022zdd,Ghoshal:2023fno} for an analysis on accounting for dark matter, matter-antimatter asymmetry, and dark radiation from PBH reheating with testable gravitational wave (GW) signatures.} and Ricci reheating \cite{Dimopoulos:2018wfg,Opferkuch:2019zbd,Bettoni:2021zhq,Figueroa:2024asq,Bettoni:2024ixe}. The Universe,  therefore, undergoes a cosmic phase of kination ($w=1$) \cite{Joyce:1997fc,Gouttenoire:2021jhk}, with the inflaton energy density getting scaled as $\rho_\phi\propto a^{-6}$, where $a$ is the FRW scale factor, before the commencement of the Hot Big Bang phase. Nonetheless, $w$ could, in principle, take any values (as long as $w>-\frac13$) in the most generic post-inflationary cosmic scenario (see e.g. Refs.~\cite{Dimopoulos:2022mce,Chen:2024roo}).

In this article, we investigate the impact of having such a non-standard post-inflationary cosmological epoch, characterized by an arbitrary $w$, on the primordial magnetic field spectrum generated during inflation. As we will see, unlike the well-known primordial GW (PGW) signals which get modified due to the background equation of state~\cite{Gouttenoire:2021jhk,Ghoshal:2022ruy}~\footnote{For instance, during a cosmic era driven by a stiff fluid with barotropic parameter $w$ lying in the range \mbox{$1/3<w<1$}, the PGW amplitude gets boosted, being detectable with future GW observatories like LISA and ET~\cite{Tashiro:2003qp, Bernal:2020ywq,Ghoshal:2022ruy,Berbig:2023yyy,Barman:2023ktz,Mishra:2021wkm,Bernal:2019lpc}.},the primordial magnetic field spectrum is insensitive to the post-inflationary cosmic history, namely the barotropic parameter ($w$) and the gauge coupling function of
the post-inflationary era.

\medskip

\section{The vector field dynamics}

We introduce below the Lagrangian of a massless $U(1)$ vector boson field reading as~\cite{Talebian:2021dfq}
\beq\label{eq:L_vector_field}
\mathcal{L}_{A} = -\frac{1}{4}I(t)\left[F_{\mu\nu}F^{\mu\nu}+\frac{\gamma}{2}F_{\mu\nu}\tilde{F}^{\mu\nu}\right],
\eeq
where $\gamma$ is a constant and $I$ is the gauge coupling function which is considered here as time-dependent. $F_{\mu\nu}$ is the Faraday tensor of the Abelian vector gauge field  $A_\mu$ written in terms of $A_\mu$ as  $F_{\mu\nu}\equiv \partial_\mu A_\nu - \partial_\nu A_\mu$. $\tilde{F}_{\mu\nu}$ is its Hodge dual, which is defined as $\tilde{F}_{\mu\nu} \equiv \frac{1}{2} \epsilon_{\mu \nu \alpha \beta} F^{\alpha \beta} $, with $\epsilon_{\mu \nu \alpha \beta}$ being the fully antisymmetric Levi-Civita symbol. Consequently, the action of our setup will read as 
\beq\label{eq:S_tot}
\begin{split}
S = & \frac{1}{16\pi G} \int d^4x \sqrt{-g}R + \int d^4x \sqrt{-g} \mathcal{L}_\mathrm{m} \\ & - \frac{1}{4}\int d^4x \sqrt{-g} I(t)\left[F_{\mu\nu}F^{\mu\nu}+\frac{\gamma}{2}F_{\mu\nu}\tilde{F}^{\mu\nu}\right],
\end{split}
\eeq
where $\mathcal{L}_\mathrm{m}$ stands for the Lagrangian density of the rest of the matter sector which does not depend on the gauge vector field $A_\mu$.

In order to quantize the gauge vector field $A_\mu$, we introduce the quantum
gauge field operator, decomposed over the full set of creation (annihilation) operators \mbox{$\hat b^\dagger_{{\bf k},\lambda} (\hat b_{{\bf k},\lambda})$} of the modes with momentum {\bf k} and transverse/circular polarization $\lambda$, as
follows
\begin{equation}
\begin{split}
\hat{\bf A}(t,{\bf x}) & =\int \frac{{\rm d}k^3}{(2\pi)^{3/2}}
\sum_{\lambda=\pm}\Biggl\{
{\bf e}_\lambda ({\bf k})\hat b_{{\bf k},\lambda}
A_\lambda(t,{\bf k}) e^{i{\bf k\cdot x}} \\ & +{\bf e}^*_\lambda ({\bf k})\hat b^\dagger_{{\bf k},\lambda}
A^*_\lambda(t,{\bf k}) e^{-i{\bf k\cdot x}}\Biggr\},
\label{Ahat}
\end{split}
\end{equation}
where the polarization three-vectors ${\bf e}_\lambda({\bf k})$ satisfy the following algebra
\begin{eqnarray}
{\bf k\cdot e}_\lambda({\bf k})=0\,,
&& {\bf e}^*_\lambda ({\bf k})={\bf e}_{-\lambda}({\bf k})\,,\nonumber\\
i{\bf k\times e}_\lambda({\bf k})=\lambda k\,{\bf e}_\lambda ({\bf k})\,, &&
{\bf e^*}_\lambda({\bf k})\cdot
{\bf e}_{\lambda'}({\bf k})=
\delta_{\lambda\lambda'}\,,
\end{eqnarray}
where \mbox{$k=|{\bf k}|$}.
Similarly, the creation and annihilation operators satisfy the canonical commutation relations
\begin{equation}
    [\hat b_{\lambda,{\bf k}},\hat b^\dagger_{\lambda',{\bf k'}}]=\delta_{\lambda\lambda'}\,
    \delta^{(3)}({\bf k}-{\bf k'})\,.
    \label{commutator}
\end{equation}

By varying the action in Eq.~\eqref{eq:S_tot} with respect to $F_{\mu\nu}$ one gets the equation of motion for the gauge field $A_\mu$ which reads as
\beq\label{eq:EOM_A_lambda_cosmic_time}
\Ddot{A}_\lambda(t,\textbf{k}) + \left(H+\frac{\dot{I}}{I}\right)\dot{A}_\lambda(t,\textbf{k}) + \left(\frac{k^2}{a^2}-\lambda\gamma\frac{k}{a}\frac{\dot{I}}{I}\right)A_\lambda(t,\textbf{k}) = 0,
\eeq
where the dot denotes a derivative with respect to the cosmic time, $\lambda = \pm$, the two helicity states of the vector gauge field, and $H$ is the Hubble parameter. Then, one can introduce the new variable $Z_\lambda(t,\textbf{k})$ defined as 
\begin{equation}
    Z_\lambda(t,\textbf{k})\equiv \sqrt{2kI}A_\lambda(t,\textbf{k})\,.
    \label{Bdef}
\end{equation}
Working with the conformal time $\eta$ defined as $\mathrm{d}\eta\equiv \mathrm{d}t/a$, \Eq{eq:EOM_A_lambda_cosmic_time} thus takes the following form:
\beq\label{eq:EOM_B_lambda_conformal_time}
\frac{\mathrm{d}^2Z_\lambda(\eta,\textbf{k})}{\mathrm{d}\eta^2}+\left(k^2 - \frac{1}{\sqrt{I}}\frac{\mathrm{d}^2\sqrt{I}}{\mathrm{d}\eta^2} -\frac{\lambda\gamma k}{I}\frac{\mathrm{d}I}{\mathrm{d}\eta}\right)Z_\lambda(\eta,\textbf{k}) = 0,
\eeq
which is a harmonic oscillator differential equation with a spacetime-dependent frequency. In the subhorizon regime, the first term in the parentheses is the dominant one and thus one recovers the Bunch-Davies solution, reading as
\beq\label{eq:Bunch_Davies_vacuum}
Z_\lambda(\eta,\textbf{k}) = e^{-ik\eta},\quad \mathrm{for}\quad k|\eta| \gg 1.
\eeq

In what follows, we will work within the context of quasi de Sitter slow-roll (SR) inflation during which the barotropic parameter $w$, defined as the ratio between the pressure $p$ and the energy density $\rho$, i.e. $w\equiv p/\rho$, is close to $-1$, 
and the slow-roll parameter $\epsilon$, defined as $\epsilon\equiv-\dot{H}/H^2$, is very small compared to unity, i.e. $\epsilon\ll 1$. Then, one can easily obtain that the Hubble parameter $H$ and the scale factor $a$ during SR inflation can be recast as
\beq
H \propto \eta^\epsilon,\quad a\propto \eta^{-1-\epsilon}.
\eeq
Then, we assume a post-inflationary era during which the barotropic parameter of the Universe is $w$ with $w$ being a free parameter varying within the range $0<w<1$ with the Hubble parameter and the  scale factor reading as
\beq
H \propto \frac{1}{\eta},\quad a\propto \eta^\frac{2}{1+3w}.
\eeq

With regard now to the gauge coupling function $I$, one should assume, in principle, a time-decreasing $I$~\cite{Sharma:2017eps,BazrafshanMoghaddam:2017zgx,Sharma:2018kgs,Sobol:2018djj,Kobayashi:2019uqs} so as not to deal with strong coupling issues~\cite{Kobayashi:2014zza}. In the following, we will consider that $I$ has the following phenomenological form:
\beq\label{eq_gauge_coupling_functions}
I = 
\begin{cases}
    I_\mathrm{ini}\left(\frac{a}{a_\mathrm{ini}}\right)^{-n_{1}},\;\mathrm{for}\; a_\mathrm{ini}<a<a_\mathrm{inf} \\
    I_\mathrm{inf}\left(\frac{a}{a_\mathrm{inf}}\right)^{-n_{2}} ,\;\mathrm{for}\; a_\mathrm{inf}<a<a_\mathrm{reh},
\end{cases}
\eeq
where $n_1$ and $n_2$ should be, in principle, positive, and the subscript `inf' denotes the end of inflation. However, as recently noticed in Ref.~\cite{Demozzi:2009fu}, a very rapid decrease of the gauge coupling function $I$ may lead to a strong backreaction. Then, one should be careful with the choice of $n_1$ and $n_2$. 
Following the analysis in Ref.~\cite{Talebian:2021dfq}, one can show that $n_1$ is bounded to the range $1<n_1\leq 4$ to avoid backreaction problems. Concerning $n_2$, we will be agnostic on its particular value.


\subsection{\boldmath The dynamics of $A_\mu$ during inflation}

Considering the parametrization of $I$, as in \Eq{eq_gauge_coupling_functions}, the equation of motion for $Z_\mu$ during inflation can be recast as

\beq\label{eq:EOM_B_conformal_time}
\frac{\mathrm{d}^2Z_\lambda(\eta,\textbf{k})}{\mathrm{d}\eta^2}+\left[k^2 - \lambda n_1\gamma\frac{k}{\eta}-\frac{n_1(n_1-2)}{4\eta^2}\right]Z_\lambda(\eta,\textbf{k}) = 0.
\eeq
The above equation can be solved analytically by imposing the Bunch-Davies initial conditions in Eq.~\eqref{eq:Bunch_Davies_vacuum} with its solution reading as

\beq
Z^{\mathrm{inf}}_\lambda(\eta,\textbf{k}) = e^{- \frac{\lambda \pi n_1\gamma}{4}}W_{\frac{i\lambda n_1\gamma}{2},\frac{n_1-1}{2}}\left(2ik\eta\right),
\eeq
where $W_{\mu,\nu}(x)$ stands for the Whittaker function of the first kind.


\subsection{\boldmath The dynamics of $A_\mu$ during the post-inflationary~era}

In order to now analytically solve \Eq{eq:EOM_B_lambda_conformal_time} during the post-inflationary era, we use the scale factor
as our time variable. In doing so, \Eq{eq:EOM_B_lambda_conformal_time} takes the following form:
\begin{widetext}
\beq\label{eq:EOM_B_lambda_scale_factor}
\frac{\mathrm{d}^2Z_\lambda(a,\textbf{k})}{\mathrm{d}a^2}+\frac{1}{a}\left(2 - \frac{3}{2}(1+w)\right)\frac{\mathrm{d}Z_\lambda(a,\textbf{k})}{\mathrm{d}a} 
+ \frac{1}{a^2}\left[\frac{k^2}{a^2H^2} - \frac{n_2(1+n_2+3w)}{4} - \lambda \gamma n_2 \frac{k}{aH}\right]Z_\lambda(a,\textbf{k}) = 0.
\eeq
\end{widetext}
Working in superhorizon scales~\footnote{

We focus here on large superhorizon scales since horizon size or subhorizon scales at the onset of the radiation-dominated era, reentering the horizon during the post-inflationary era, are really small compared with the scales accessible by current CMB and large-scale structure probes. Only scales which are superhorizon during the post-%
inflationary era crossing the horizon during the radiation- or the matter-dominated eras can be observationally probed.
}, i.e. $k\ll aH$, one can neglect the vacuum term proportional to $k^2$ in \Eq{eq:EOM_B_lambda_scale_factor}.  Equation~\eqref{eq:EOM_B_lambda_scale_factor} is then recast as
\begin{widetext}
\beq\label{eq:EOM_B_lambda_scale_factor_2}
\frac{\mathrm{d}^2Z_\lambda(a,\textbf{k})}{\mathrm{d}a^2}+\frac{1}{a}\left[2 - \frac{3}{2}(1+w)\right]\frac{\mathrm{d}Z_\lambda(a,\textbf{k})}{\mathrm{d}a} 
- \frac{1}{a^2}\left[ \frac{n_2(1+n_2+3w)}{4} +\lambda \gamma n_2 \frac{k}{aH}\right]Z_\lambda(a,\textbf{k}) = 0.
\eeq
\end{widetext}
Equation~\eqref{eq:EOM_B_lambda_scale_factor_2} accepts an analytic solution, reading as 
\begin{widetext}
\beq
\begin{split}\label{eq:B_post_inflation}
Z^\mathrm{post-inf}_\lambda(\eta,\textbf{k}) & = \frac{2}{1+3w}(-1)^{-\frac{n_2}{1+3w}}\sqrt{\frac{k}{a_\mathrm{inf}H_\mathrm{inf}}}\left(\frac{a(\eta)}{a_\mathrm{inf}}\right)^{\frac{1+3w}{4}} \\ & \times \Biggl\{ c_{1,w} I_{-1-\frac{2n_2}{1+3w}}\left(\frac{4}{1+3w}\sqrt{\lambda n_2\gamma}\sqrt{\frac{k}{a_\mathrm{inf}H_\mathrm{inf}}}\left(\frac{a(\eta)}{a_\mathrm{inf}}\right)^{\frac{1+3w}{4}}\right) \Gamma\left(-\frac{2n_2}{1+3w}\right) \\ & -(-1)^{\frac{2n_2}{1+3w}} c_{2,w} I_{1+\frac{2n_2}{1+3w}}\left(\frac{4}{1+3w}\sqrt{\lambda n_2\gamma}\sqrt{\frac{k}{a_\mathrm{inf}H_\mathrm{inf}}}\left(\frac{a(\eta)}{a_\mathrm{inf}}\right)^{\frac{1+3w}{4}}\right) \Gamma\left(2 +\frac{2n_2}{1+3w}\right)\Biggr\},
\end{split}
\eeq
\end{widetext}
where $I_n(z)$ is the modified Bessel function of the first kind and $c_{1,w}$ and $c_{2,w}$ are integration constants depending on the comoving scale $k$, while $a(\eta) = a_\mathrm{inf}\left(\frac{\eta}{\eta_\mathrm{inf}}\right)^{\frac{2}{1+3w}}$ during the post-inflationary era. The integration constants $c_{1,w}$ and $c_{2,w}$ will be determined by imposing continuity of the vector field $A_\lambda(\eta,\textbf{k})$ 
and its first time derivative at the end of inflation. In particular, one should impose that
\begin{align}
A^\mathrm{inf}_\lambda(\eta_\mathrm{inf},\textbf{k}) &  = A^\mathrm{post-inf}_\lambda(\eta_\mathrm{inf},\textbf{k}) \\ (A'_\lambda)^\mathrm{inf}(\eta_\mathrm{inf},\textbf{k}) & = (A'_\lambda)^\mathrm{post-inf}(\eta_\mathrm{inf},\textbf{k}).
\end{align}
After a straightforward but lengthy calculation, one gets that
\begin{widetext}
\beq
\begin{split}
c_{1,w} & = - \frac{(-1)^{\frac{n_2}{1 + 3 w}}}{8\lambda n_2\gamma}(1+w)^2
      e^{-\frac{\lambda n_1 \pi \gamma}{4}}
   \Biggl[-\frac{4}{1+3w}
     \sqrt{\frac{k}{H_\mathrm{inf}a_\mathrm{inf}}}\sqrt{\lambda n_2\gamma} I_{\frac{2 n_2}{1 + 3 w}}\left(\frac{4}{1+3w}\sqrt{\lambda n_2\gamma}\sqrt{\frac{k}{a_\mathrm{inf}H_\mathrm{inf}}}\right) W_{\frac{i\lambda n_1\gamma}{2}, \frac{n_1-2}{2}}\left(-\frac{2ik}{a_\mathrm{inf}H_\mathrm{inf}}
     \right) \\ & + I_{1+\frac{2 n_2}{1 + 3 w}}\left(\frac{4}{1+3w}\sqrt{\lambda n_2\gamma}\sqrt{\frac{k}{a_\mathrm{inf}H_\mathrm{inf}}}\right) \\ & \times \Bigl\{i\left[-\frac{2k}{a_\mathrm{inf}H_\mathrm{inf}} + n_1(i-\lambda\gamma)\right]W_{\frac{i\lambda n_1\gamma}{2}, \frac{n_1-2}{2}}\left(-\frac{2ik}{a_\mathrm{inf}H_\mathrm{inf}}\right) -W_{1+\frac{i\lambda n_1\gamma}{2}, \frac{n_1-2}{2}}\left(-\frac{2ik}{a_\mathrm{inf}H_\mathrm{inf}}\right) \Bigr\} \Biggl]
         \\ & \mbox{\Huge /}\Biggl\{\left(\frac{k}{a_\mathrm{inf}H_\mathrm{inf}}\right)\Gamma\left(-\frac{2n_2}{1+3w}\right)\Biggl[I_{\frac{2 n_2}{1 + 3 w}}\left(\frac{4}{1+3w}\sqrt{\lambda n_2\gamma}\sqrt{\frac{k}{a_\mathrm{inf}H_\mathrm{inf}}}\right)I_{-1-\frac{2 n_2}{1 + 3 w}}\left(\frac{4}{1+3w}\sqrt{\lambda n_2\gamma}\sqrt{\frac{k}{a_\mathrm{inf}H_\mathrm{inf}}}\right) \\ & - I_{\frac{-2 n_2}{1 + 3 w}}\left(\frac{4}{1+3w}\sqrt{\lambda n_2\gamma}\sqrt{\frac{k}{a_\mathrm{inf}H_\mathrm{inf}}}\right)I_{1+\frac{2 n_2}{1 + 3 w}}\left(\frac{4}{1+3w}\sqrt{\lambda n_2\gamma}\sqrt{\frac{k}{a_\mathrm{inf}H_\mathrm{inf}}}\right)\Biggr]\Biggr\}
\end{split}
\eeq

\beq
\begin{split}
c_{2,w} & = \frac{(-1)^{\frac{-n_2}{1 + 3 w}}}{\Gamma\left(2+\frac{2n_2}{1+3w}\right)}e^{-\frac{\lambda n_1 \pi \gamma}{4}}
      \frac{1 + 3 w}{8(\lambda n_2)^{3/2}\gamma}\frac{a_\mathrm{inf}H_\mathrm{inf}}{k} \\ & \times \Biggl[4\sqrt{\frac{k}{a_\mathrm{inf}H_\mathrm{inf}}} \lambda n_2\sqrt{\gamma} 
        I_{-\frac{2 n_2}{1 + 3 w}}\left(\frac{4}{1+3w}\sqrt{\lambda n_2\gamma}\sqrt{\frac{k}{a_\mathrm{inf}H_\mathrm{inf}}}\right) W_{\frac{i\lambda n_1\gamma}{2}, \frac{n_1-2}{2}}\left(-\frac{2ik}{a_\mathrm{inf}H_\mathrm{inf}}\right)
        \\ &  + \sqrt{\lambda n_2}\left[\left(\frac{2ik}{a_\mathrm{inf}H_\mathrm{inf}} + i\lambda n_1\gamma + n_1\right)(1+3w) \right] I_{-1-\frac{2 n_2}{1 + 3 w}}\left(\frac{4}{1+3w}\sqrt{\lambda n_2\gamma}\sqrt{\frac{k}{a_\mathrm{inf}H_\mathrm{inf}}}\right) W_{\frac{i\lambda n_1\gamma}{2}, \frac{n_1-2}{2}}\left(-\frac{2ik}{a_\mathrm{inf}H_\mathrm{inf}}\right)
        \\ & + 2\sqrt{\lambda n_2}\left(1 + 3w\right)I_{-1-\frac{2 n_2}{1 + 3 w}}\left(\frac{4}{1+3w}\sqrt{\lambda n_2\gamma}\sqrt{\frac{k}{a_\mathrm{inf}H_\mathrm{inf}}}\right) W_{1+\frac{i\lambda n_1\gamma}{2}, \frac{n_1-2}{2}}\left(-\frac{2ik}{a_\mathrm{inf}H_\mathrm{inf}}\right)\Biggr]
        \\ & \mbox{\Huge /}\Biggl[I_{\frac{2 n_2}{1 + 3 w}}\left(\frac{4}{1+3w}\sqrt{\lambda n_2\gamma}\sqrt{\frac{k}{a_\mathrm{inf}H_\mathrm{inf}}}\right)I_{-1-\frac{2 n_2}{1 + 3 w}}\left(\frac{4}{1+3w}\sqrt{\lambda n_2\gamma}\sqrt{\frac{k}{a_\mathrm{inf}H_\mathrm{inf}}}\right) 
         \\ & - I_{-\frac{2 n_2}{1 + 3 w}}\left(\frac{4}{1+3w}\sqrt{\lambda n_2\gamma}\sqrt{\frac{k}{a_\mathrm{inf}H_\mathrm{inf}}}\right)I_{1+\frac{2 n_2}{1 + 3 w}}\left(\frac{4}{1+3w}\sqrt{\lambda n_2\gamma}\sqrt{\frac{k}{a_\mathrm{inf}H_\mathrm{inf}}}\right) \Biggr] 
\end{split}
\eeq

\end{widetext}

\section{The primordial magnetic field power spectrum}

Since we are interested in superhorizon modes which reenter the cosmological horizon during the post-inflationary era, we perform an expansion of the integration constants $c_{1,w}$ and $c_{2,w}$ and the modified Bessel functions present in \Eq{eq:B_post_inflation} in the superhorizon regime, i.e. $k\ll aH$. In doing so, one gets

\begin{align}\label{eq:c_1_c2_approx}
 c_{1,w}(k\ll aH) & \propto
\begin{cases}
k^{1-\frac{n_1}{2}}k^{\frac{n_2}{1+3w}},\;\mathrm{for}\;n_1>1 \\
k^{\frac{n_1}{2}}k^{\frac{n_2}{1+3w}},\;\mathrm{for}\;n_1<1,
\end{cases}
\\
 c_{2,w}(k\ll aH) & \propto
\begin{cases}
k^{-\frac{n_1}{2}}k^{-\frac{n_2}{1+3w}},\;\mathrm{for}\;n_1>0 \\
k^{\frac{n_1}{2}}k^{-\frac{n_2}{1+3w}},\;\mathrm{for}\;n_1<0.
\end{cases}
\end{align}

\begin{align}\label{eq:I_approx}
I_{-1-\frac{2n_2}{1+3w}}\left(k\ll aH\right) & \propto  k^{-\frac{n_2}{1+3w}-\frac{1}{2}}\\
I_{1+\frac{2n_2}{1+3w}}\left(k\ll aH\right) & \propto   k^{\frac{n_2}{1+3w}+\frac{1}{2}}.
\end{align}

By plugging  \Eq{eq:c_1_c2_approx} and \Eq{eq:I_approx} into \Eq{eq:B_post_inflation}, we obtain 
\beq
\begin{split}
& Z^\mathrm{post-inf}_\lambda(\eta,k\ll aH)  = \frac{2}{1+3w}(-1)^{-\frac{n_2}{1+3w}}\sqrt{\frac{k}{a_\mathrm{inf}H_\mathrm{inf}}} \\ & \times \left(\frac{a(\eta)}{a_\mathrm{inf}}\right)^{\frac{1+3w}{4}}\Biggl\{ c_{1,w} I_{-1-\frac{2n_2}{1+3w}}\left(k\ll aH\right) \Gamma\left(-\frac{2n_2}{1+3w}\right) \\ & -(-1)^{\frac{2n_2}{1+3w}} c_{2,w} I_{1+\frac{2n_2}{1+3w}}\left(k\ll aH\right) \Gamma\left(2 +\frac{2n_2}{1+3w}\right)\Biggr\}.
\end{split}
\eeq 

Thus, with regard to the spectral shape of $Z^\mathrm{post-inf}_\lambda(\eta,k)$ on superhorizon scales, one gets 
\beq
\begin{split}
Z^\mathrm{post-inf}_\lambda(\eta,k\ll aH) & \propto c_{1,w}\left(k\ll aH\right) I_{-1-\frac{2n_2}{1+3w}}\left(k\ll aH\right) \\ & + c_{2,w}\left(k\ll aH\right) I_{1+\frac{2n_2}{1+3w}}\left(k\ll aH\right)
\\ & \propto k^{\frac{1-|1-n_1|}{2}},
\end{split}
\eeq
while the Fourier mode of the gauge boson field $A_\lambda(\eta,\textbf{k})$ can be recast as
\beq\label{eq:A_superhorizon}
A_\lambda(\eta,k\ll aH) = \frac{Z_\lambda(\eta,k\ll aH)}{\sqrt{2kI}} \propto
k^{-\frac{|1-n_1|}{2}}.
\eeq

Then, one can ultimately compute the magnetic field power spectrum defined as~\cite{Subramanian:2009fu}
\beq
P_B(\eta,k)\equiv \frac{\mathrm{d}\mathrm{\rho}_B}{\mathrm{d}\ln k} = \sum_{\lambda=\pm}\frac{k^5}{4\pi^2a^4}I|A_\lambda(\eta,k)|^2.
\eeq
Finally, in the case of the presence of both non-minimal kinetic and axial couplings of the form \eqref{eq:L_vector_field} where the gauge vector field is given by \Eq{eq:A_superhorizon} on superhorizon scales, it is straightforward to see that 
\beq \label{eq:P_B_superhorizon} 
P_B(\eta,k\ll aH)\propto
k^{5-|1-n_1|}.
\eeq
Interestingly enough, as one may infer from \Eq{eq:P_B_superhorizon}, we obtain a magnetic field primordial power spectrum with a spectral shape insensitive to the barotropic parameter of the post-inflationary era $w$. We concluded this result starting with a Lagrangian of a massless $U(1)$ vector boson field of the form in Eq.~\eqref{eq:L_vector_field}, which is a generic class of electromagnetic Lagrangians with non-minimal kinetic and axial couplings. We also checked that our result is robust if one considers either  kinetic or axial couplings, thus switching on or off the constant parameter~$\gamma$. We expect that our result is also valid even if the gauge kinetic and axial couplings behave differently.

At this point, we need to stress as well that the spectral shape of the  magnetic field power spectrum is sensitive only to the exponent $n_1$, i.e. on the dynamical evolution of the gauge coupling function $I$ during inflation, and does not depend on the exponent $n_2$, i.e. behaviour of $I$ during the post-inflarionary era~\footnote{
The independence of the magnetic field power spectrum from the power-law index $n_2$, although not explicitly stated, has also been found in Ref.~\cite{Sharma:2018kgs} but only for the case of a matter-dominated post-inflationary era. In this work, we performed a more generic analysis finding both
a $w$ and $n_2$ independent large-scale magnetic field power spectrum in the post-inflationary
era.
}. As a consistency check, we need to mention that one can easily check that for $n_1 = 4$ we recover from \Eq{eq:P_B_superhorizon} the scale-invariant attractor solution $P_B\propto k^2$~\cite{Dimopoulos:2012av}.

In our analysis, we have implicitly assumed that our gauge field and its associated magnetic field do not interact with any post-inflation plasma before reheating. However, in many reheating mechanisms, there is already a thermal bath present before reheating, which is subdominant to the energy density of the Universe but may not be subdominant to the primordial magnetic field. In this case, magnetohydrodynamic effects cannot be ignored. If important, such effects would conserve magnetic flux for an Abelian gauge field with a canonical kinetic coupling only. Non-zero magnetic field helicity would result in inverse cascade, which would transfer power to larger scales. For non-canonical kinetic couplings, however, their time dependence may have profound implications on the magnetic field spectrum (see, for example, Refs.~\cite{Brandenburg:2021pdv,Brandenburg:2021bfx}). We feel that this possibility is beyond the scope of this letter. In effect, we assume that the thermal bath is not present before reheating, as would be the case, for example, in curvaton reheating~\cite{Feng:2002nb,BuenoSanchez:2007jxm}, or in reheating due to the evaporation of primordial black holes~\cite{Lennon:2017tqq,Martin:2019nuw,Dalianis:2021dbs}.
\vspace{0.15cm}

\section{Discussion and Conclusions} 

It is intriguing that the existence of a non-standard post inflationary cosmological epoch could have shaped the morphology of the primordial magnetic field spectrum. But it turns out that the final large-scale magnetic field spectrum is quite insensitive to the barotropic parameter $w$ and to the dynamical evolution gauge coupling function during the post-inflationary era, namely, on the power-law index $n_2$.
This means that even though we may observe the impact of any post inflationary epoch via its signatures, for instance, in the CMB spectrum itself, as it impacts the duration or number of $e$-folds during reheating \cite{Martin:2010kz,Martin:2014nya,Martin:2021frd}, thereby correlating predictions for the inflationary observables, such as $n_s$ and $r$, with other associated signals, like gravitational waves - or via its effect on the spectral shape of gravitational waves \cite{Tashiro:2003qp, Bernal:2020ywq,Ghoshal:2022ruy,Berbig:2023yyy,Barman:2023ktz,Mishra:2021wkm,Bernal:2019lpc}, the magnetic field spectrum itself may not carry any such information. This result leads to the conclusion that the primordial magnetic field spectrum is a pure carrier of the microphysics of inflation rather than of the post-inflationary evolution of the Universe.

\medskip

{\bf{Acknowledgments}}--
KD is supported (in part) by the STFC consolidated Grant:
ST/X000621/1. 
TP acknowledges the contribution of the LISA Cosmology Working Group and the COST Action
CA21136 ``Addressing observational tensions in cosmology with systematics and 
fundamental physics (CosmoVerse)''. TP also acknowledges the support of INFN Sezione di Napoli \textit{iniziativa specifica} QGSKY as well as financial support from the Foundation for Education and European Culture in Greece.

\bibliography{references}

\end{document}

%% file: newcommands.tex
\let\oldsqrt\sqrt
\def\sqrt{\mathpalette\DHLhksqrt}
\def\DHLhksqrt#1#2{%
\setbox0=\hbox{$#1\oldsqrt{#2\,}$}\dimen0=\ht0
\advance\dimen0-0.2\ht0
\setbox2=\hbox{\vrule height\ht0 depth -\dimen0}%
{\box0\lower0.4pt\box2}}

\newcommand{\beq}{\begin{equation}}
\newcommand{\eeq}{\end{equation}}
\newcommand{\bea}{\begin{equation}\begin{aligned}}
\newcommand{\eea}{\end{aligned}\end{equation}}

\newlength{\wsingfig}
\newlength{\wdblefig}
\newlength{\wquadfig}
\newlength{\wtriplefig}
\newcommand{\Eq}[1]{Eq.~(\ref{#1})}